# The Pattern Speeds of M51, M83 and NGC 6946 Using CO and the Tremaine-Weinberg Method

P. Zimmer, R. J. Rand, and J. T. McGraw
Department of Physics and Astronomy, University of New Mexico,
800 Yale NE, Albuquerque, NM 87106;
zimm@as.unm.edu, rjr@as.unm.edu, mcgraw@as.unm.edu

## Abstract

In spiral galaxies where the molecular phase dominates the ISM, the molecular gas as traced by CO emission will approximately obey the continuity equation on orbital timescales. The Tremaine-Weinberg method can then be used to determine the pattern speed of such galaxies. We have applied the method to single-dish CO maps of three nearby spirals, M51, M83 and NGC 6946 to obtain estimates of their pattern speeds: 38 ± 7 km/s/kpc, 45 ± 8 km/s/kpc and 39 ± 8 km/s/kpc, respectively, and we compare these results to previous measurements. We also analyze the major sources of systematic errors in applying the Tremaine-Weinberg method to maps of CO emission.

# 1. INTRODUCTION

The pattern speed, $\Omega_P$, of density wave phenomena is one of the most fundamental parameters for understanding the inner workings of spiral and barred galaxies. Pattern speeds are not directly observable but crucial for understanding the rate by which spirals and bars affect the evolution of galaxies. Spiral density waves have been shown to trigger star formation in grand-design spirals (e.g. Rand 1993; Knapen et al. 1996), so the rate at which gas encounters spiral arms may affect the rate of star formation. Scattering of stars off spiral arms is likely to be a significant source of disk heating, which may be particularly important in late-type spirals (Merrifield, Gerssen & Kuijken 2001). The rate of heating in this way is therefore linked to the rate at which stars encounter the arms, and thus the pattern speed. Accurate pattern speeds also place strong constraints on different theories of spiral structure as well as the bar-spiral connection.

It has proven somewhat elusive to determine pattern speeds observationally, however. The most common technique involves identifying specific resonances between the spiral pattern's propagation and the orbits of stars and gas in the galaxy (e.g. Toomre 1969; Tully 1974; Elmegreen, Elmegreen & Seiden 1989; Canzian 1993). Such techniques do not always yield unambiguous results. For example, one observational difficulty is identifying the radial distance at which spiral structure ends, which depends upon the depth of the imaging and the choice of tracer (e.g. stars vs. HI). Tracer selection led to substantially different results in the case of M81, for which observable spiral structure "ends" at < 10 kpc for stars vs. > 16 kpc for HI (Westpfahl 1998). Furthermore, certain resonances may not exist in some galaxies depending on the circular velocity curve and pattern speed. A method of measuring the pattern speed

*not* based on identifying such morphological features would allow independent determination of the resonance locations (as long as circular velocity curves are known) and examination of associated morphological features, thus providing tests for different theories of spiral structure, as well as addressing the issues in the preceding paragraph. Tremaine & Weinberg (1984; hereafter TW) describe such a technique for measuring the pattern speed as a function of distance along the minor axis of barred and spiral galaxies.

The TW method is a straightforward technique which uses readily observable quantities. It requires measurement of the surface density and velocity of a component of the galaxy (stars, atomic or molecular gas, etc. -- any component that reacts to the density wave) along lines or apertures parallel to the major axis of the galaxy. The method is more thoroughly covered in Section 2, but the most crucial requirement, and most difficult to satisfy, is that the component be continuous (that is, satisfies the continuity equation) over the course of several orbits about the center of the galaxy. There must also be a well-understood tracer of the surface density of the component of the galaxy, such as 21-cm emission for HI, visible or IR light for stars and CO emission for molecular gas.

To date, the method has most successfully been applied to the stars in the bar patterns of the SB0 galaxies NGC 936 (Merrifield & Kuijken 1995), NGC 4596(Gerssen, Merrifield, & Kuijken 1999), and NGC 7079 (Debattista 1998). Because there is little star formation in the SB0 galaxies, stars meet the continuity requirement in that they are not being created or lost at a significant rate compared to the galactic orbital period. Optical light, in most disk galaxies, does not effectively trace the surface mass density of stars because of dust obscuration, hence the almost exclusive use of nearly dust-free SB0 galaxies. Sambhus & Sridhar (2000) have used a modified form of the method to analyze the pattern speed of the nucleus of M31 and Debattista, Gerhard, & Sevenster (2002) have applied another form of the TW method to OH/IR stars in the Milky Way.

In most galaxies, neither atomic hydrogen nor molecular gas will obey the continuity equation, because of conversion of gas between phases (molecular « atomic « ionized) and star formation on orbital and shorter timescales. In select galaxies where either the atomic or the molecular phase everywhere comprises the large majority of the ISM, the conversion processes can be neglected, as they will not significantly affect the surface density of the dominant phase at any location. Furthermore, as long as the star formation efficiency is low (on average, spiral disks convert ~5% of their gas into stars every $10^8$ years; Kennicutt 1998), conversion of gas into stars (and vice-versa through winds and supernovae) can be ignored. Galaxies where HI dominates the ISM can be particularly useful, as 21-cm observations readily yield both intensity and velocity. The TW method has been applied to two such galaxies: M81 (Westpfahl 1998) and the dwarf barred-spiral galaxy NGC 2915 (Bureau et al. 1999).

To the extent that $^{12}$CO(1-0) emission traces molecular gas surface density, the TW method can also be applied to galaxies with $H_2$-dominated ISMs. With this idea in mind, we obtained existing high-resolution, single-dish CO maps for M51, M83 and NGC 6946 and applied the TW method to them.

Section 2 discusses the TW method in more detail and our requirements for applying it to CO data. Section 3 describes information about each of the data sets. Section 4 describes our analysis of the CO data for the three galaxies and the resulting pattern speeds along with a comparison with previous measurements of the pattern speed of each of the galaxies. We summarize and conclude in Section 5.

## 2. THE TREMAINE-WEINBERG METHOD

TW describe a model-independent means of measuring the pattern speed of bars in SB0 galaxies. Its application requires measurement of the distribution of intensity and velocity of a component that reacts to the density wave and that satisfies a few requirements:

1) There is only one well-defined pattern speed for the galaxy.

2) The component must satisfy the continuity equation over several orbits around the galaxy, i.e. the component must not be created or destroyed in any significant amount.

3) There must be a well-understood tracer of the component's surface mass density, such that it can be inferred from the intensity of the tracer. The radial velocity of the component is also required.

4) The surface density of the component (and therefore the intensity of the tracer of that component) must go to zero at the edges of the galaxy.

5) The disk of the galaxy must be flat (no warps), at least out to a distance where the intensity of the tracer is nearly zero. Motions and variations in structure perpendicular to the disk must be negligible.

In practice, these requirements must be met for the region of the galaxy that is being analyzed.

Following TW, we will refer to a galaxy's coordinates in the following manner. We place the origin of our coordinate system at the dynamical center of the galaxy and align the *x*-axis with the major axis of the galaxy. The *y*-axis is then aligned with the minor axis of the galaxy, and for the following derivation, will be measured in the frame of the galaxy, related to sky-coordinates by a factor of *sin(i)*, where *i* is the inclination of the galaxy. Velocities are measured with respect to the systemic velocity of the galaxy.

As a consequence of the first requirement, we can write the surface density, $\Sigma$, in polar coordinates as $\Sigma(r, f, t) = \Sigma(r, f_0 - \Omega_P t)$, where the time dependence of the surface density is due only to the rotation of the pattern. The continuity equation, because of the second requirement, takes the form

$$\frac{\partial \Sigma(x, y, t)}{\partial t} + \frac{\partial}{\partial x}[\Sigma(x, y, t) v_x(x, y, t)] + \frac{\partial}{\partial y}[\Sigma(x, y, t) v_y(x, y, t)] = 0 \quad (1)$$

which is just the 2D continuity equation with no source or sink terms, and $v_x$ and $v_y$ are just the Cartesian components of the velocity field. Because $\Omega_P$ is assumed constant, we can write

$$\frac{\partial \Sigma}{\partial t} = \frac{\partial \Sigma}{\partial \phi}\frac{\partial \phi}{\partial t} = -\Omega_P \frac{\partial \Sigma}{\partial \phi} = -\Omega_P \left( x \frac{\partial \Sigma}{\partial y} - y \frac{\partial \Sigma}{\partial x} \right) \quad (2)$$

and we have converted from polar coordinates to Cartesian. We then combine Equations 1 and 2 and solve for $\Omega_P$:

$$\Omega_P = \frac{\frac{\partial}{\partial x}(\Sigma v_x) + \frac{\partial}{\partial y}(\Sigma v_y)}{x\frac{\partial \Sigma}{\partial y} - y\frac{\partial \Sigma}{\partial x}}. \tag{3}$$

Westpfahl (1998) notes that, if one knows the derivatives of $\Sigma$, $v_x$, and $v_y$, (and all the above requirements are satisfied) one can solve for $\Omega_P$ point by point using Equation 3. Unfortunately, given our definition of coordinate frame, $v_x$ is the proper motion of the galaxy component, which is generally not observable in external galaxies. Also, the derivative of a measured quantity is inherently noisier than the quantity itself.

TW show that the terms involving $x$-derivatives can be eliminated by integrating the numerator and denominator from $x = -\infty$ to $x = +\infty$, incorporating Requirement 4, yielding:

$$\Omega_P = \frac{\int_{-\infty}^{+\infty} \frac{\partial}{\partial y}(\Sigma v_y) dx}{\int_{-\infty}^{+\infty} x \frac{\partial \Sigma}{\partial y} dx}. \tag{4}$$

This integration effectively averages $\Omega_P$ over lines (henceforth apertures) parallel to the major axis of the galaxy. The numerator and denominator still depend on the $y$-derivative of the surface density and velocity, which, though calculable from 2D intensity and velocity maps, are not directly observable, and are noisy as mentioned above. Therefore, TW integrate the numerator and denominator again, this time parallel to the minor axis, eliminating the $y$-derivatives but maintaining the proportionality and casting the relationship in terms of readily observable quantities (we have asserted Requirement 3, that $I(x) \propto \Sigma(x)$, and that $v_y sin(i) = v_{LOS}$,

with $v_{LOS}$ measured with respect to the systemic velocity of the galaxy):

$$\Omega_P = \frac{\int_{-\infty}^{+\infty} \Sigma v_y \, dx}{\int_{-\infty}^{+\infty} x \Sigma \, dx} = \frac{1}{\sin(i)} \frac{\int_{-\infty}^{+\infty} I(x) v_{LOS}(x) \, dx}{\int_{-\infty}^{+\infty} x I(x) \, dx} \tag{5}$$

Merrifield & Kuijken (1995) refine this technique. They normalize the numerator and denominator with the total intensity in an aperture, such that Equation 5 becomes

$$\Omega_P = \frac{1}{\sin(i)} \frac{\int_{-\infty}^{+\infty} I(x) v_{LOS}(x) \, dx}{\int_{-\infty}^{+\infty} x I(x) \, dx} \frac{\int_{-\infty}^{+\infty} I(x) \, dx}{\int_{-\infty}^{+\infty} I(x) \, dx} = \frac{1}{\sin(i)} \frac{<V(x)>}{<x>}. \tag{6}$$

The numerator becomes the intensity-weighted line of sight velocity and the denominator the intensity-weighted position of the tracer along the aperture in question. These quantities can be easily extracted from longslit spectra, or, for data cubes, from maps of total intensity and velocity (zeroth and first moment maps). One then measures these values for a number of apertures in a given galaxy, plots $<V>$ vs. $<x>$ and fits a line, the slope of which is the average pattern speed of the galaxy (times $sin(i)$). This refinement has the benefit of both evening out the statistical importance of each aperture-- because apertures closer to the major axis of the galaxy have both greater intensities and higher projected velocities--as well as avoiding a singularity where the denominator goes to zero near the center of the galaxy (Bureau, Freeman, & Pfitzner 1999). It also eliminates adverse effects in the original TW method due to errors in the dynamical center and systemic velocity estimates for the galaxy analyzed.

The numerator in the TW integral measures a surface mass flux across each aperture, as

*I(x)* is assumed to be proportional to the surface density. The denominator measures where the bulk of the tracer is crossing the aperture, with respect to the minor axis of the galaxy, which is related to the tangential gradient of the surface mass density. For a pattern moving with constant angular velocity, this ratio should be constant. For an axisymmetric, rotating disk with no pattern, every aperture will be symmetric in intensity and anti-symmetric in velocity about the minor axis (*x*=0) giving a null result (zero over zero). There will be no net flux because as much material flows through the aperture in one direction on one side of the minor axis as in the other direction on the other side. An overdensity of the tracer superimposed on a smooth disk will generate a non-zero pattern speed for apertures that cross it, which is why one is driven to use more than one aperture. A density wave pattern will cover more than one aperture, and hopefully many. The contributions of random density fluctuations over many apertures should average to zero, but those from a rotating density wave pattern should add. If the overdensities associated with the pattern are not sufficiently large as to overwhelm random fluctuations in the tracer density, this technique will give inconsistent results, and so strong patterns yield more robust results.

CO emission is generally considered a reasonable tracer of molecular gas surface density, and for the most part, we will assume this to be so and that the conversion factor between $H_2$ column density and CO intensity is constant. However, we shall test this hypothesis by analyzing a few alternatives in the next section. A particular choice of CO/$H_2$ conversion factor will not affect our result, because it cancels out of the TW integral so long as it is constant over a given galaxy (although extreme choices could invalidate our assumption of $H_2$-dominance).

## 3. THE GALAXIES AND THE DATA

We require maps made with a single dish telescope because interferometric maps without zero-spacing data will miss large-scale flux (violating the continuity requirement). The need to clearly resolve the spiral structure of the galaxy drives the need for maps of nearby galaxies made with telescopes of diameter 30m and larger. Several such maps exist in the literature for galaxies with molecule-dominated ISMs, and we have obtained maps of three different galaxies: M51, M83 and NGC6946, all in $^{12}$CO(1-0).

### 3.1 M51

M51 is an excellent candidate for application of the TW method due to its angular size and the strength of its arms. As well, its ISM is clearly molecule dominated (García-Burillo, Guélin, & Cernicharo 1993, Nakai et al. 1994). The global $H_2$/HI mass ratio is 1.8 using the $H_2$ mass of Scoville & Young (1983) and the HI mass of Rots(1980). More importantly, Rand, Kulkarni, & Rice (1992) show that in radial profiles, the surface density of $H_2$ (from García-Burillo et al. 1993) is nowhere less than three times that of HI (from Rots et al. 1990) within 7 kpc from the center (using a CO®$H_2$ conversion factor of $N(H_2)/I_{CO} = 3 \times 10^{20}$ mol cm$^{-2}$ (K km/s)$^{-1}$). The star formation rate is 3.78 $M_\odot$/yr (Kennicutt et al. 94, Kennicutt 1998), corresponding to a Roberts time (gas consumption time) of 3.98 Gyr (not including any recycling factor), which is considerably larger than the orbital timescales in the region mapped. Hence, consumption of $H_2$ by star formation is not a significant sink in the continuity equation.

Although M51 is an interacting system, the interior region is thought to sustain a den-sity wave which may be driven by an outer, material spiral pattern (e.g. Tully 1974; Elmegreen et al. 1989; Howard and Byrd 1990; Salo and Laurikainen 2000). Even if the strong spiral structure of M51 is due to the interaction with NGC5195, as long as a density wave is present and the aforementioned conditions are met, the TW method still applies. M51 also has a small central bar (Pierce 1986; Zaritsky & Lo 1986).

We acquired an existing map of M51 obtained by Nakai et al. (1994) with the 45-m Nobeyama Telescope in $^{12}$CO(1-0), kindly provided by N. Nakai. The map covers 6' x 6' with 16" resolution sampled at 3" per pixel. For the purposes of our analysis, we adopt a distance of 9.5 Mpc, an inclination of 20º, major axis position angle of -10º and kinematic center of RA $13^h27^m46^s$, Dec. 47º27'10" (B1950) as given in Nakai et al. (1994 and references therein). The map was provided as a data cube, from which we derived 3s-clipped zeroth- and first-moment maps.

### 3.2 M83

Our $^{12}$CO(1-0) map of M83 was obtained and provided by A. Andersson et al. (2001) with the 30m SEST telescope. The data were provided as zeroth- and first-moment maps and cover 9' x 9' with 22" resolution sampled at 11" per pixel. We assume an inclination of 24º, a major axis position angle of 45º, a distance of 5.0 Mpc (Lord & Kenney 1991) and a kinematic center of RA $13^h34^m12^s$, Dec. -29º36'38" (J2000).

The ISM of M83 is also clearly dominated by molecular gas in the regions where the CO flux is significant. Crosthwaite et al. (2002) shows that the ratio of molecular to atomic hydrogen density is at least 3 to a radius of 6.5 kpc (using a CO→H$_2$ conversion factor of N(H$_2$)/I$_{CO}$ = 3 x 10$^{20}$ Mol. cm$^{-2}$ (K km/s)$^{-1}$). In addition, M83 has a very strong bar, and even though it may have a different pattern speed than the arms, the strength of its spiral pattern makes it another good candidate. The star formation rate is 2.56 M$_☉$/yr (Kennicutt et al. 94, Kennicutt 1998) which corresponds to a Roberts time of 3.87 Gyr.

### 3.3 NGC6946

W. Walsh kindly furnished a map of NGC 6946 taken with the 30m IRAM telescope (Walsh et al. 2002). The data were provided as zeroth- and first-moment maps and cover 10' x 10' with 21" resolution sampled at 10" per pixel. We adopt the inclination angle of 38º, position angle of 240º, distance of 5.5 Mpc and kinematic center of RA 20$^h$34$^m$52.7$^s$, Dec 60º09'13" (J2000) from Walsh et al. (2002).

NGC 6946 should be an excellent test of the versatility of the TW method as it has a relatively weak four-armed pattern. Tacconi & Young (1990) found the azimuthally averaged H$_2$/HI ratio to be greater than 3 (using a CO→H$_2$ conversion factor of N(H$_2$)/I$_{CO}$ = 2.8 x 10$^{20}$ Mol. cm$^{-2}$ (K km/s)$^{-1}$) within a radius of 2.25' (almost all of the detectable CO in the first-moment map is within 3' of the center) and so its ISM qualifies as molecule dominated in the inner regions for the purposes of our analysis. The star formation rate for NGC 6946 is 1.14 M$_☉$/yr (Kennicutt 94, Kennicutt et al. 1998) which corresponds to a Roberts time of 14.0 Gyr.

## 4. RESULTS

Each map was analyzed using a MATLAB program designed to extract one pixel wide apertures parallel to the major axis, spaced by half of a beamwidth, from each of the intensity and velocity maps. A second program then calculated $<V>$ and $<x>$ for each aperture. We then plot $<V>$ vs. $<x>$ and fit a line to the values, the slope of which, divided by $\sin(i)$, gives the derived pattern speed. We estimate the errors in our values by combining the inherent uncertainty in the method given by TW of 15% with those due to the uncertainty in the position angle of the major axis (see section 4.4). Though there are many possible sources of systematic errors, we test the most important of these using the Nobeyama map of M51. A summary of all our results is shown in Table 1.

### 4.1 M51

In Figure 1, we present the Nobeyama map of M51 with the apertures we extracted superimposed on it. Figure 2 shows $<V>$ vs. $<x>$ for the Nobeyama M51 map along with the fit value of the pattern speed of 38 ± 7 km/s/kpc. The data exhibit a well-defined slope. We have excluded apertures at the edge of the map because they give inconsistent results due to the weakness of the CO at the edges of the map. Beyond 6.5 kpc, they may also be contaminated by the outer spiral pattern, which is thought to be a material spiral pattern with a lower pattern speed, driven by the encounter with NCG 5195 (e.g. Tully 1974). The apparent change in the pattern speed in the center of the galaxy may be because the responsible apertures cross M51's small bar, which may be moving with a pattern speed different from the spiral arms. This apparent change may also be an artifact of the highly asymmetrical CO emission in the central region if the asymmetry is not due to the density wave pattern. If this change in derived pattern speed

Table 1:

Pattern Speed Determinations

| M51 | $\Omega_P$ (km/s/kpc) |
|---|---|
| All Apertures | $38 \pm 7$ |
| Bar | $> 100$ |
| Radially Enhanced CO$\rightarrow$H$_2$ | $38 \pm 7$ |
| Enhanced Interarm CO$\rightarrow$H$_2$ | $42 \pm 7$ |
| Odd-weighting | $32 \pm 7$ |
| Previous Measurements | $40^1$ |
| Modeled Values | $27^2, 50^3$ |
| M83 | |
| All Apertures | $45 \pm 8$ |
| Previous Measurements | $51^4$ |
| NGC 6946 | |
| All Apertures | $39 \pm 8$ |
| Odd-weighting | $34 \pm 7$ |

1) Tully (1974); Elmegreen et al. (1989).

2) García-Burillo et al. (1993).

3) Salo and Laurikainen (2000).

4) Lord & Kenney (1991).

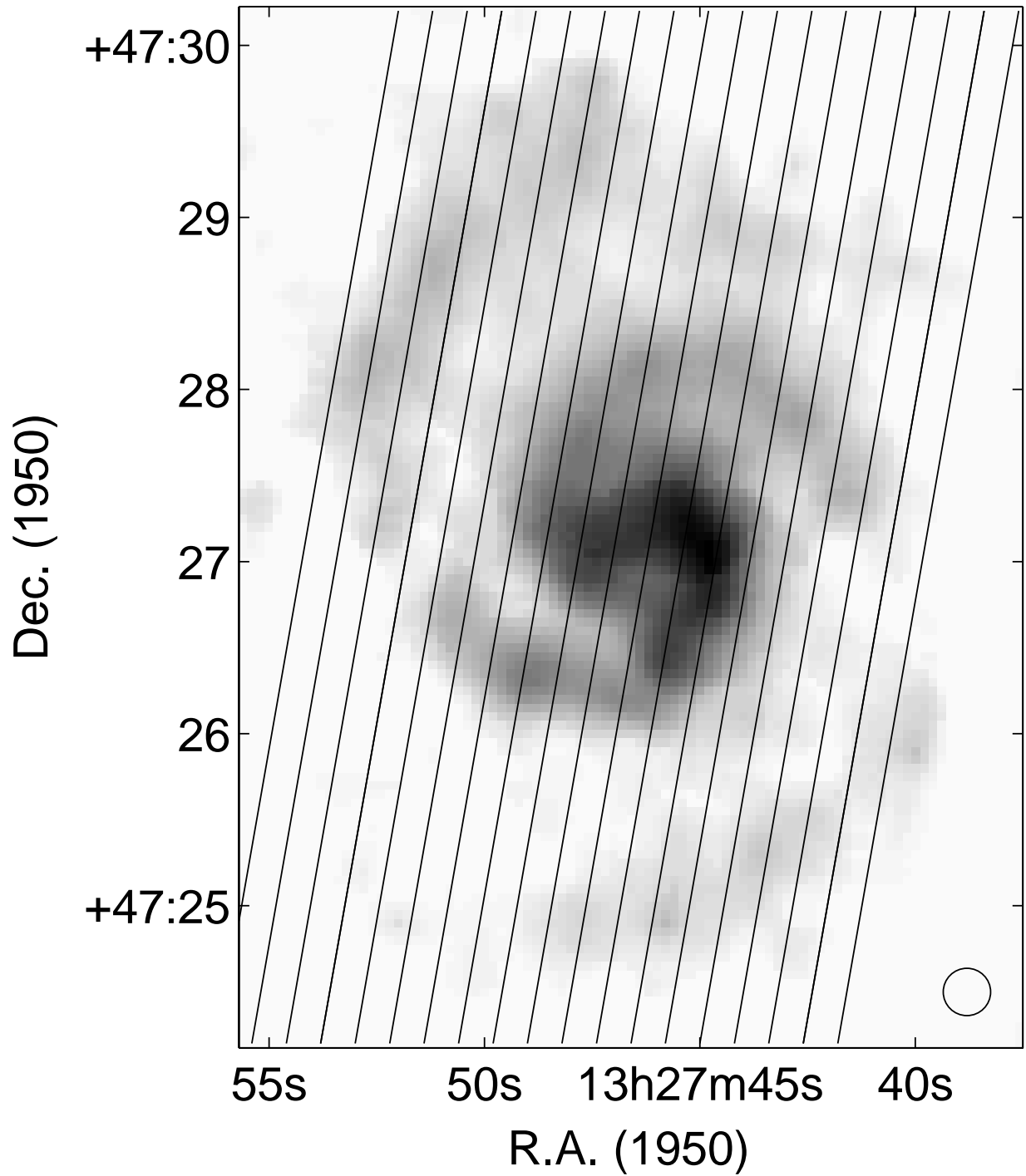

Figure 1: Grayscale map of $^{12}$CO(1-0) emission in M51 measured by Nakai et al. (1994). The virtual apertures (black lines) used in the TW method analysis are superimposed. Only every other aperture is shown for clarity.

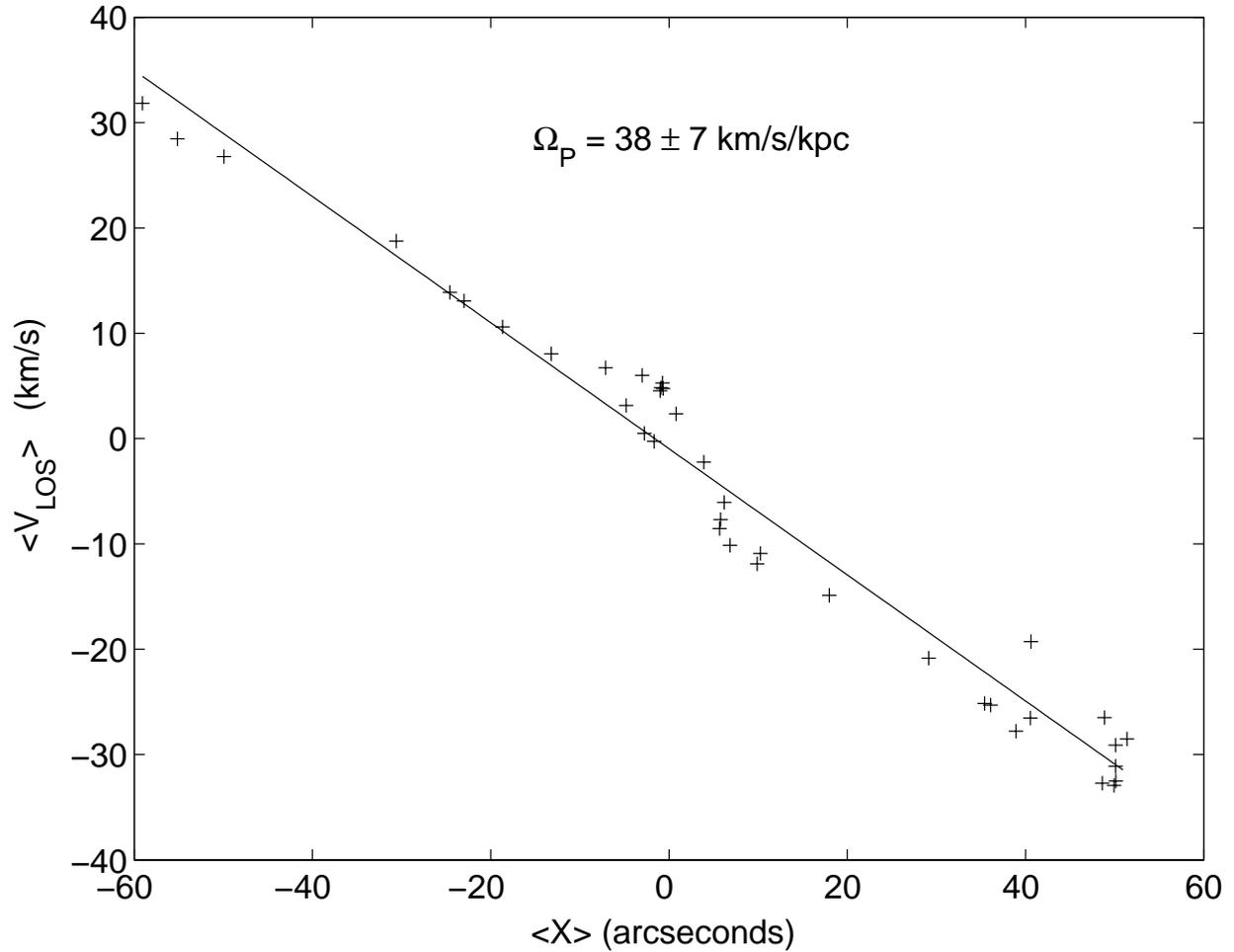

Figure 2: $\langle V \rangle$ vs. $\langle x \rangle$ for each aperture from the map of M51 in Figure 1 are plotted and a straight line is fit, the slope of which is the pattern speed.

is due to a different pattern speed in the bar, we can set a lower limit on the bar's pattern speed of $\Omega_P \geq 100$ km/s/kpc, because apertures containing both the bar and the spiral pattern will have a mixture of the two speeds. However, exclusion of these potentially contaminated apertures does not significantly affect the derived pattern speed.

For comparison, the generally accepted value of the inner pattern speed of M51, when corrected for our adopted distance, is that of Tully (1974) of 40 km/s/kpc, which, given the errors inherent in each value, is in good agreement with our value of 38 ± 7 km/s/kpc. Tully identified the inner Lindblad resonance and corotation as the inner and outer terminations of the spiral pattern as traced by Hα emission, respectively. Elmegreen et al. (1989) identified several resonances in M51, most notably the 4:1 resonance which corresponds to a minimum in starlight along the spiral arm, and derived a pattern speed consistent with that of Tully and in agreement with our estimate. However, García-Burillo, Combes, & Gerin (1993) simulated the dynamics of molecular clouds in M51 to determine the pattern speed. By comparing the modeled and observed gas morphology, they found the most likely pattern speed to be significantly lower, 27 km/s/kpc. The tidal interaction model of Salo and Laurikainen (2000) requires a pattern speed of ~50 km/s/kpc in the inner parts of M51 so that there is no Inner Lindblad Resonance and spiral waves can propagate into the central 30", explaining observed K-band features. However, a strong feature of their model is that waves increase in pattern speed as they propagate toward the center, a situation for which the TW method is inappropriate.

**4.2 M83**

In Figure 3 we plot the positions of our apertures on the map of M83 and in Figure 4 we show $<V>$ vs. $<x>$ and the fit value of $\Omega_P$ of 45 ± 8 km/s/kpc, which includes errors of ± 4 km/s/kpc from the uncertainty in the position angle. Once again, the slope is well-defined. One can see that there is no evidence for a change in pattern speed induced by the very large and strong bar of M83. This may be due to the orientation of the bar, parallel to the major axis, which will tend to minimize any signal in the TW method (as it is sensitive to asymmetries in the intensity and velocity about the minor axis). Alternatively, the bar may have the same pattern speed as the spiral arms and may be driving the strong, two-armed spiral pattern. In support of this posibility are the findings of Elmegreen & Elmegreen (1995), that the bar length in barred spirals correlates with the size of the two-armed domain of disks.

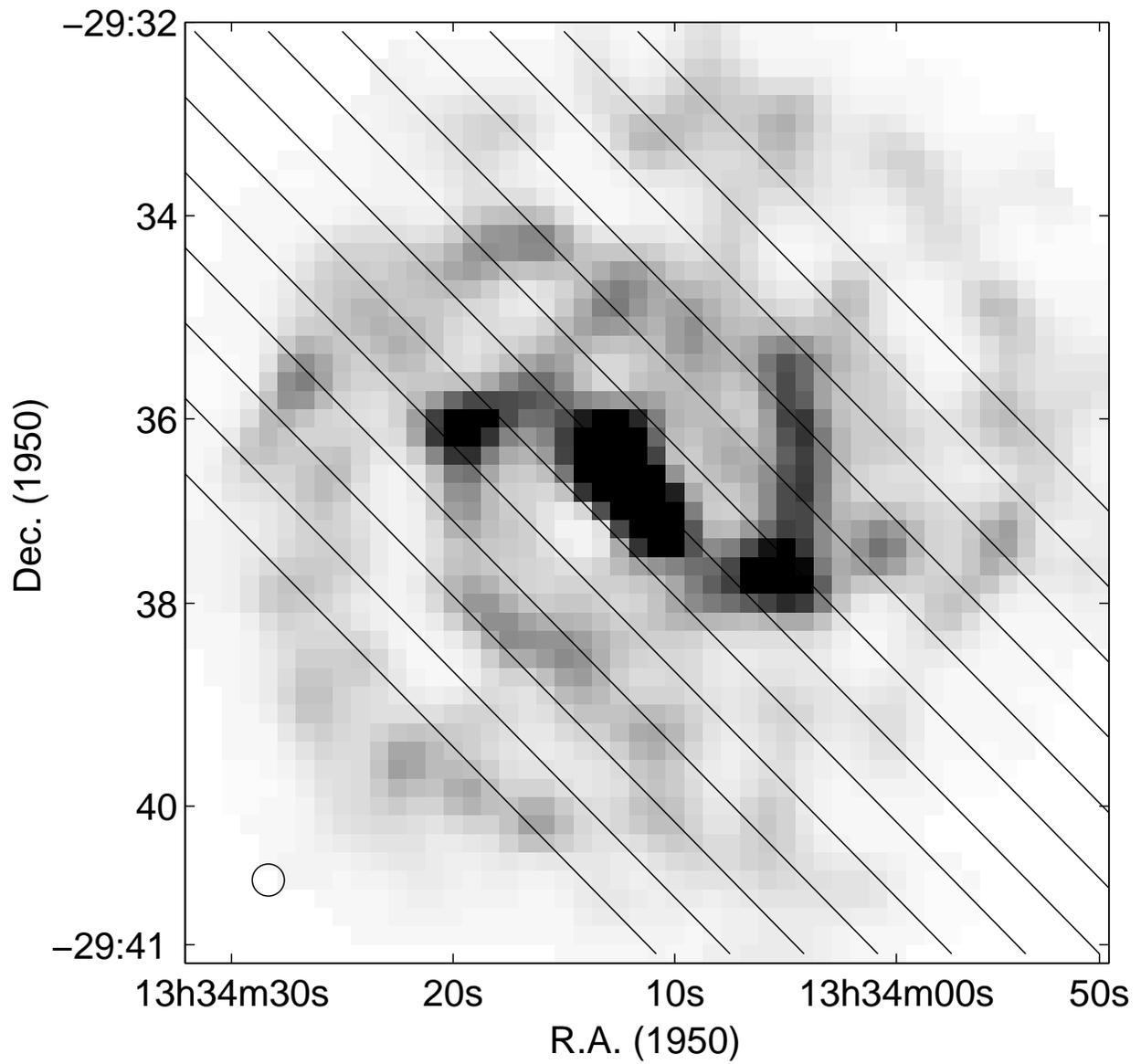

Figure 3: CO map of M83 with virtual apertures. The grayscale plot is $^{12}$CO(1-0) emission as measured by Anderson et al. (2001), with every fourth aperture shown.

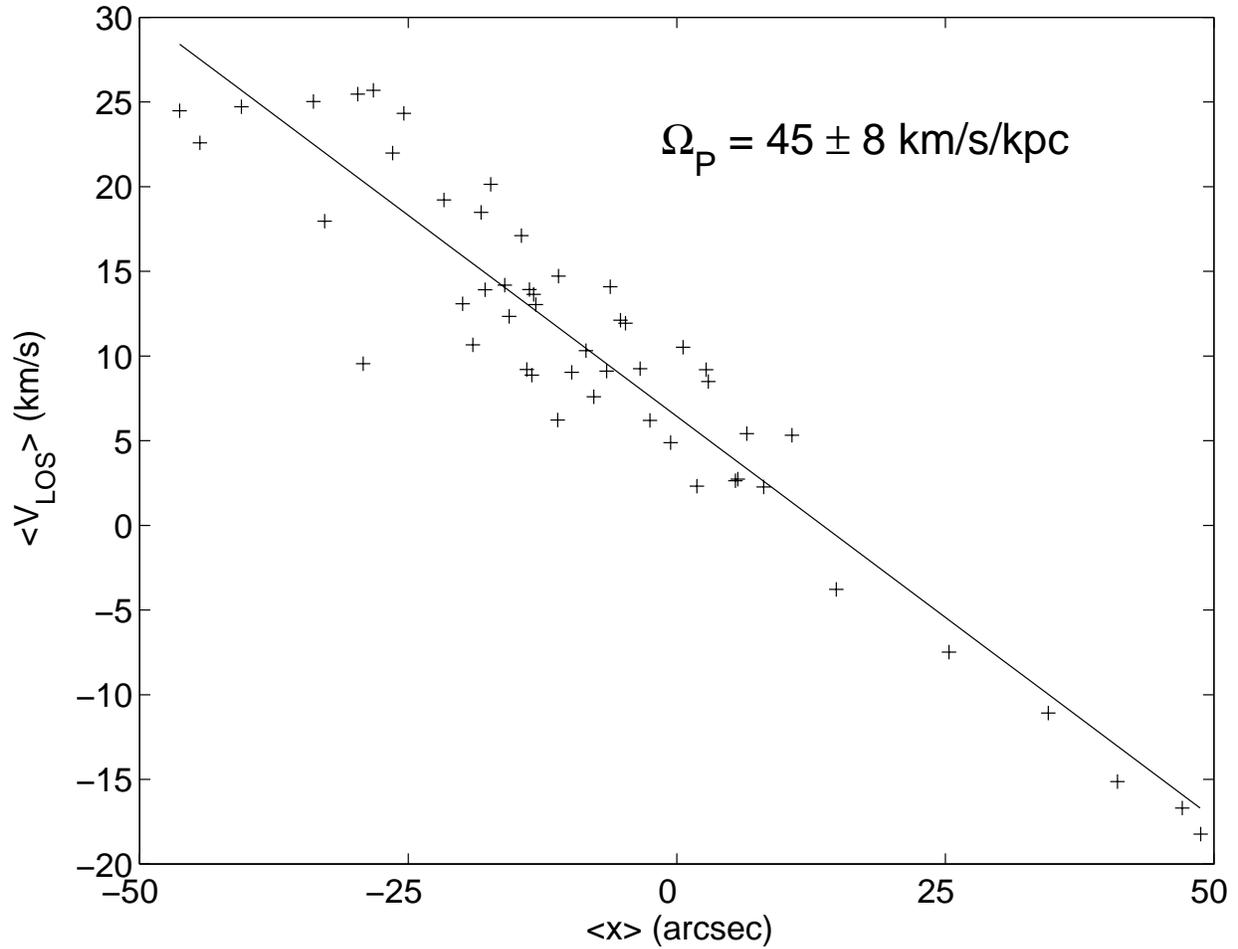

Figure 4: $<V>$ vs. $<x>$ for each aperture from the map of M83 in Figure 3 are plotted and a straight line is fit, the slope of which is the pattern speed.

Lord & Kenney (1991) measured the pattern speed of M83 by identifying the corotation resonance as the location of where the HII regions and dust lanes cross the stellar arm, and found a value of 51 km/s/kpc. Again, this value is remarkably close to the value of 45 ± 8 km/

s/kpc we derive.

### 4.3 NGC 6946

Figure 5 plots the positions of the apertures for the map of NGC 6946 and Figure 6 plots $\langle V \rangle$ vs. $\langle x \rangle$. As in M83, a single well-defined slope is found. Our resulting value of $\Omega_P$ is 39 $\pm$ 8 km/s/kpc, including errors $\pm 5$ km/s/kpc from the uncertainty in the position angle of the ma-jor axis. We know of no previously published values for the pattern speed of NGC 6946.

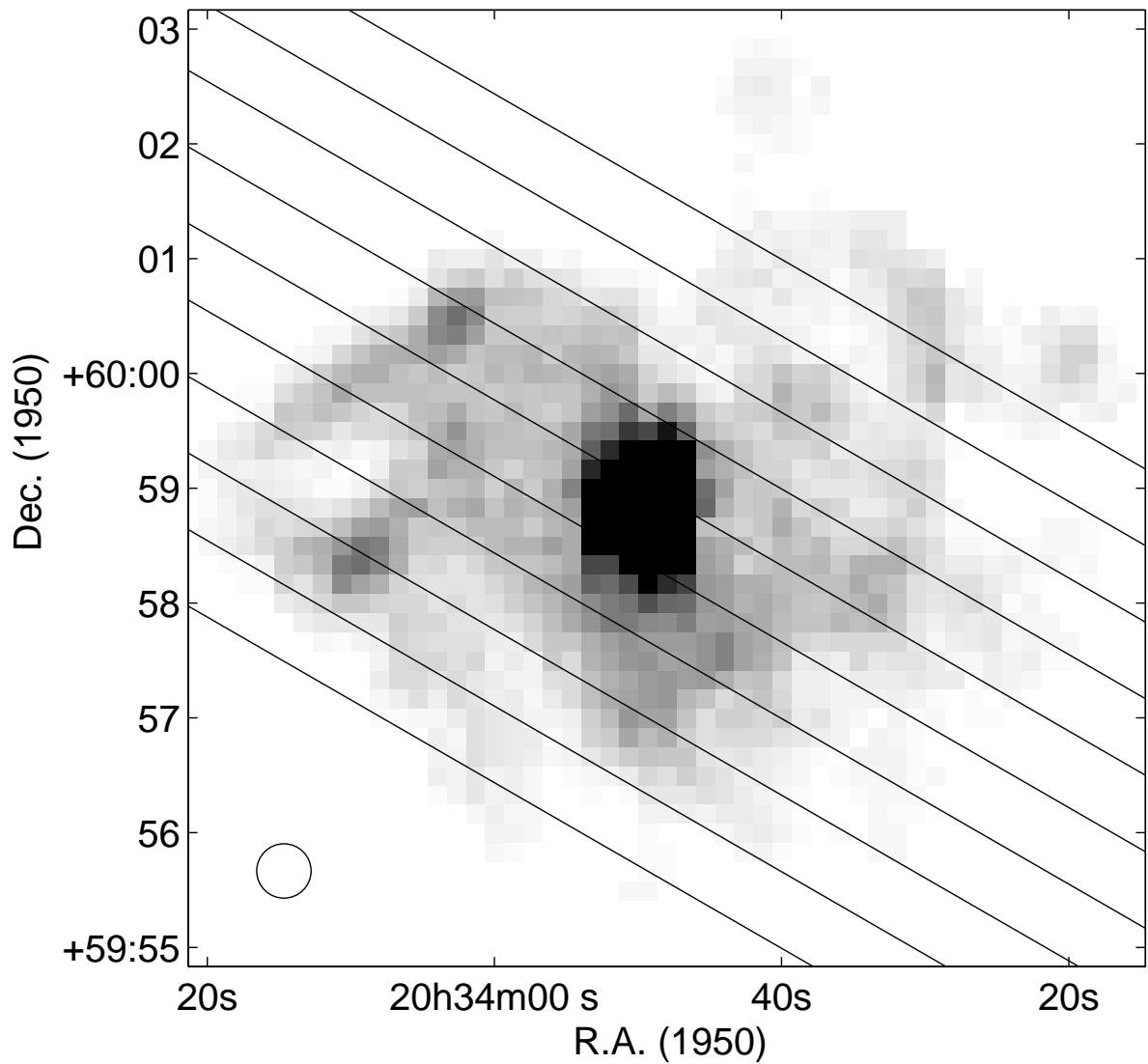

Figure 5: CO map of NGC 6946 with virtual apertures. The grayscale plot is $^{12}$CO(1-0) emission as measured by Walsh et al. (2002) with every fourth aperture shown.

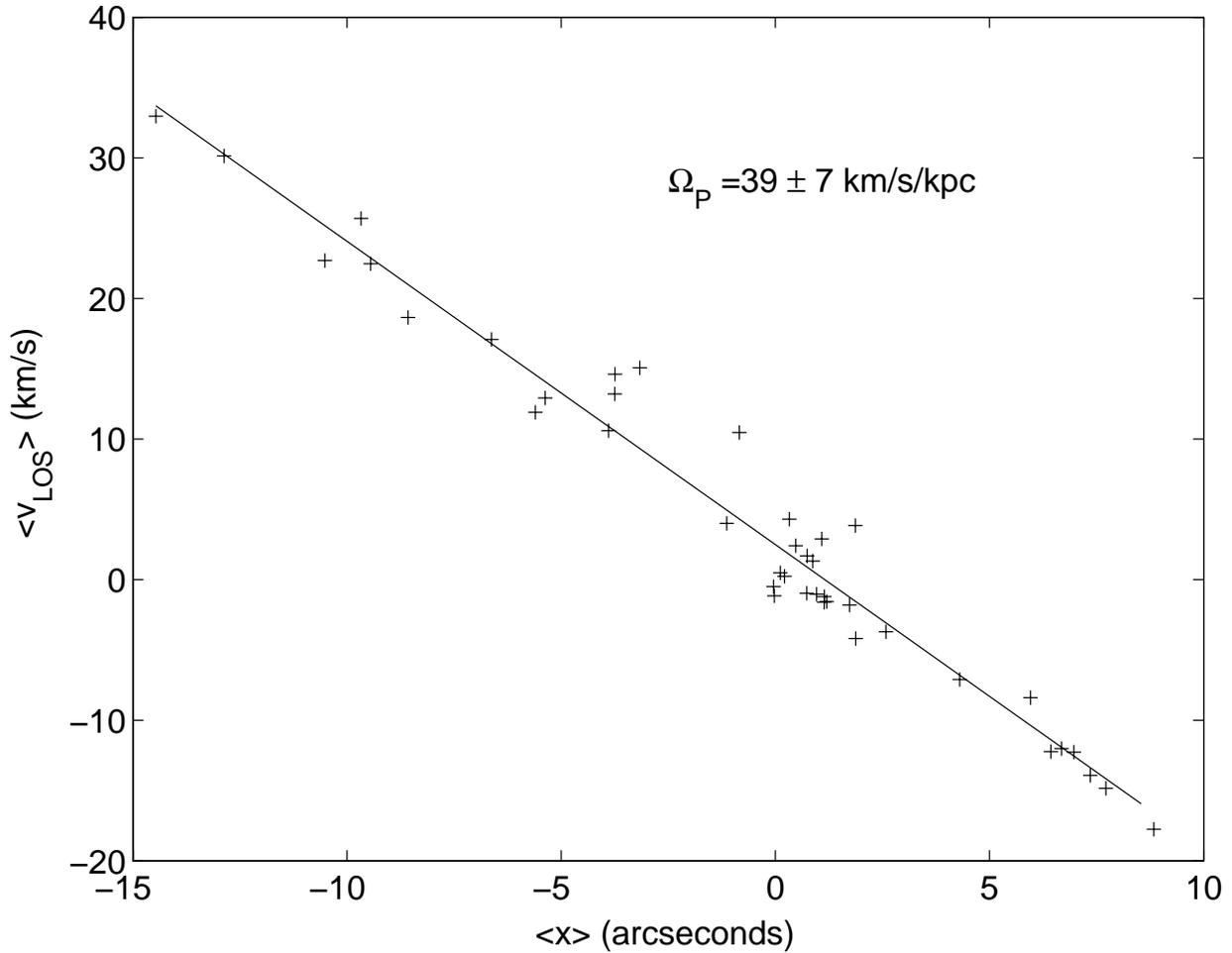

Figure 6: $\langle V \rangle$ vs. $\langle x \rangle$ for each aperture from the map of NGC 6946 in Figure 5 are plotted and a straight line is fit, the slope of which is the pattern speed.

### 4.4 TESTS OF POTENTIAL SYSTEMATIC EFFECTS

We conducted tests of several likely sources of systematic error using the Nobeyama map. The TW method is rather sensitive to different choices of the position angle of the major axis of the galaxy. We test the effect of varying the position angle and examine the resulting change in the derived pattern speed. Through the generally accepted range of values $-10° \pm 5°$ for M51 (Tully 1974), the resulting pattern speed varies by ± 3 km/s/kpc, which is included in our error bars. Larger offsets (of 15° or more) have dramatic results and tend to destroy any linear relation between <V> and <x>. Sensitivity to map resolution was tested by boxcar smoothing both the intensity and velocity maps by factors of two and three. The difference in resulting pattern speed is less than 1 km/s/kpc in both cases. Smoothing by much larger amounts, though, gives ambiguous results – the number of independent apertures becomes small and the scatter increases, so that the resulting error bars for the fitted pattern speed become large.

We also examined the effects of a spatially variable value of the CO®$H_2$ conversion factor. It is possible that this factor may have a radial dependence due to a metallicity gradient, because a lower metallicity may lead to a lower abundance of CO (Arimoto, Sofue, & Tsujimoto 1996; Wilson 1995; Boselli, Lequeux, & Gavazzi 2002; Dumke et al. 1997; Lequeux 1996 and references therein). Similarly, an arm-interarm contrast of the conversion factor may exist due to heating, which drives the conversion factor in the arm regions down, or macroscopic opacity effects, which drives the conversion factor in the arms up (García-Burillo et al. 1993, but c.f. Rand 1993). We therefore made naïve modifications to the intensity map of M51 as a simple way of simulating the heating effect by dividing the intensity map into arm and interarm regions and enhancing the interarm intensities by a factor of two to test the case where heating may have enhanced the arm emission by this factor. To test for some radial dependence of the conversion factor in M51, we simply applied an intensity correction, which varies in radius inversely with the known metallicity gradient of 0.04 dex/kpc (Pagel & Edmunds 1981), following the galaxy-to-galaxy scaling of Arimoto et al. (1996). This produces a threefold en-hancement of the CO intensity at the edges of the apertures. In each case, the effect on the derived pattern speed was minimal: for the radial enhancement $\Omega_P = 38 \pm 7$ km/s/kpc, and for the enhanced interarm emission $\Omega_P = 42 \pm 7$ km/s/kpc. We do not expect the radial enhancement to affect our derived pattern speeds significantly for modest gradients as the radial enhancement enters as a zero-mean weighting factor for both $<V>$ and $<x>$, maintaining their proportionality. Moreover, it matters less for apertures further from the major axis, which carry more weight in the fit, because the fractional radius variation along them is smaller. Tests of radial enhancement factor for M83 and NGC 6946 using the measured metallicity gradients from Vila-Costas and Edmunds (1992) also yielded no significant change in computed pattern speed.

It is also possible that these maps do not cover a large enough spatial extent or are not sufficiently deep to account for the bulk of the CO emission. If this is the case, then the integral used to calculate $<V>$ and $<x>$ may not have converged even for integrations extending to the edge of the map. To test this possibility, we plot in Figure 7, for each aperture in M51, values of $<V>$ and $<x>$ in which we grow the limits of integrations from the minor axis to where the aperture reaches the two edges of the map. Nearly half of the apertures appear to have not fully converged to a stable value by the edge of the map, although most of the variation of these integrals occurs for limits of integration less than 100" from the major axis. Figure 8 plots $<V>$ divided by $<x>$ for M51 as a function of distance from the minor axis, where we have adjusted the axes to show the behavior at the edge of the map, and it appears that roughly a quarter of the apertures have not converged to a stable $<V>/<x>$. The large variation in final $<V>/<x>$ values is due to apertures near the major axis where $<x>$ is nearly zero. This result simply highlights one of the advantages of determining the pattern speed from plots of $<V>$ vs. $<x>$. Similar plots for M83 and NGC 6946 show similar behavior. This may indicate that the maps used for our analysis are not deep enough and the assumptions behind the TW method do not hold over the mapped region. However, in Figures 9-11, we show the best fit pattern speed for M51, M83 and NGC 6946 if the integration is stopped at six different distances from the minor axis. The derived pattern speeds are well within our error budget for all three galaxies for limits of inte-gration halfway to the edge of the map or greater. Even though some of the individual apertures may not have converged to well-defined values of <V>, <x>, or <V>/<x> for a few apertures near the major axis, the best-fit value of the pattern speed for all the apertures has converged well before the limits of integration reach the edge of the map. These tests do indicate that the depth of the maps used can be an important factor and the above tests are a good way of judging whether or not a given map is sufficiently deep to successfully apply the TW method.

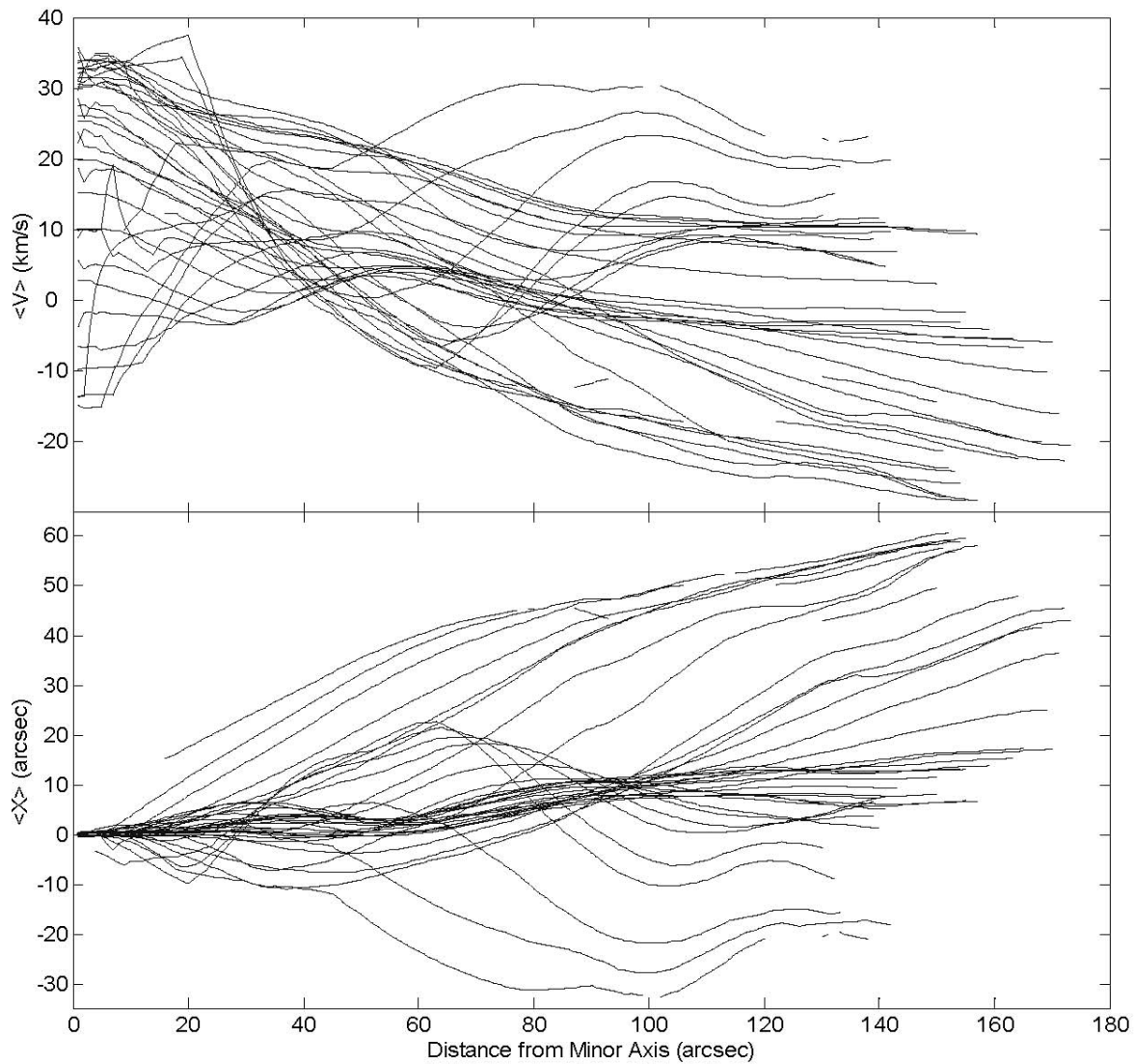

Figure 7: Values of <V> and <x> for which the limits of integration, shown along the horizontal axis, are increased from zero to where the aperture reaches the two edges of the map.

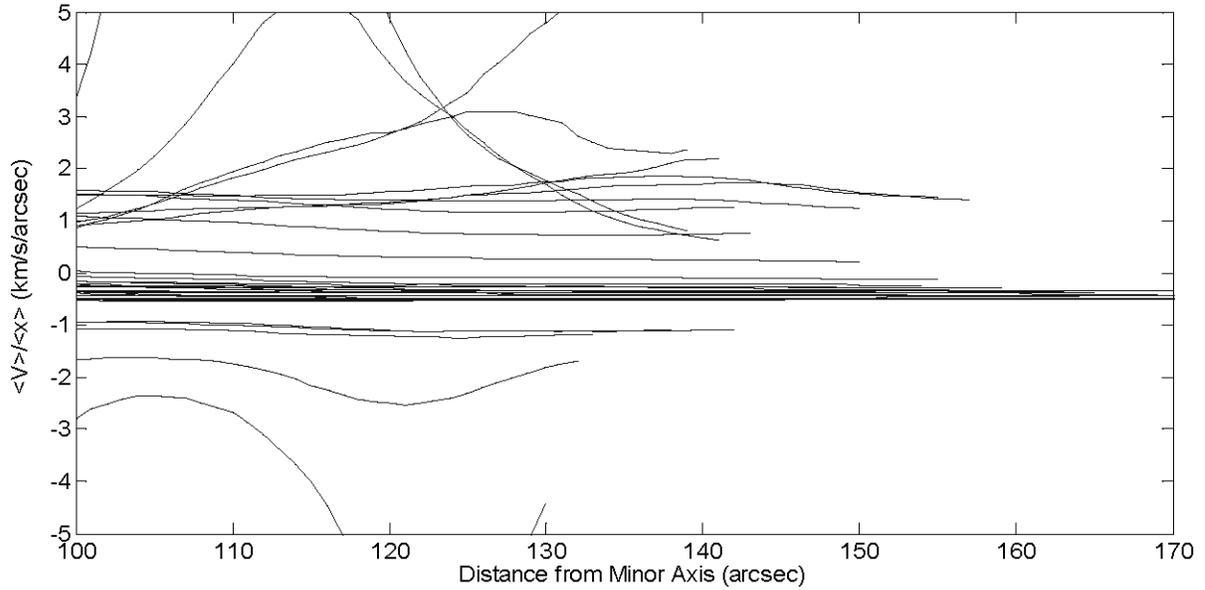

Figure 8: Values of <V>/<x> for which the limits of integration, shown along the horizontal axis, are increased from zero to where the aperture reaches the two edges of the map.

Finally, from looking at the intensity maps of M51 and NGC 6946, we see that they differ significantly from bisymmetry. This may be due to odd spiral modes which may propagate with a different pattern speed than that of the even modes. TW show that applying an odd weighting function to their original technique eliminates the effects of odd spiral modes. There-fore we have adapted their idea by combining symmetrically spaced pairs of apertures on either side of the major axis such that $<V_{odd}>=<V_+>-<V_->$ and $<x_{odd}>=<x_+>-<x_->$ and plotting $<V_{odd}>$ vs. $<x_{odd}>$. For both galaxies, the resulting pattern speed is lower than the unweighted case, but not significantly: $32 \pm 7$ km/s/kpc for M51 and $34 \pm 7$ km/s/kpc for NGC 6946.

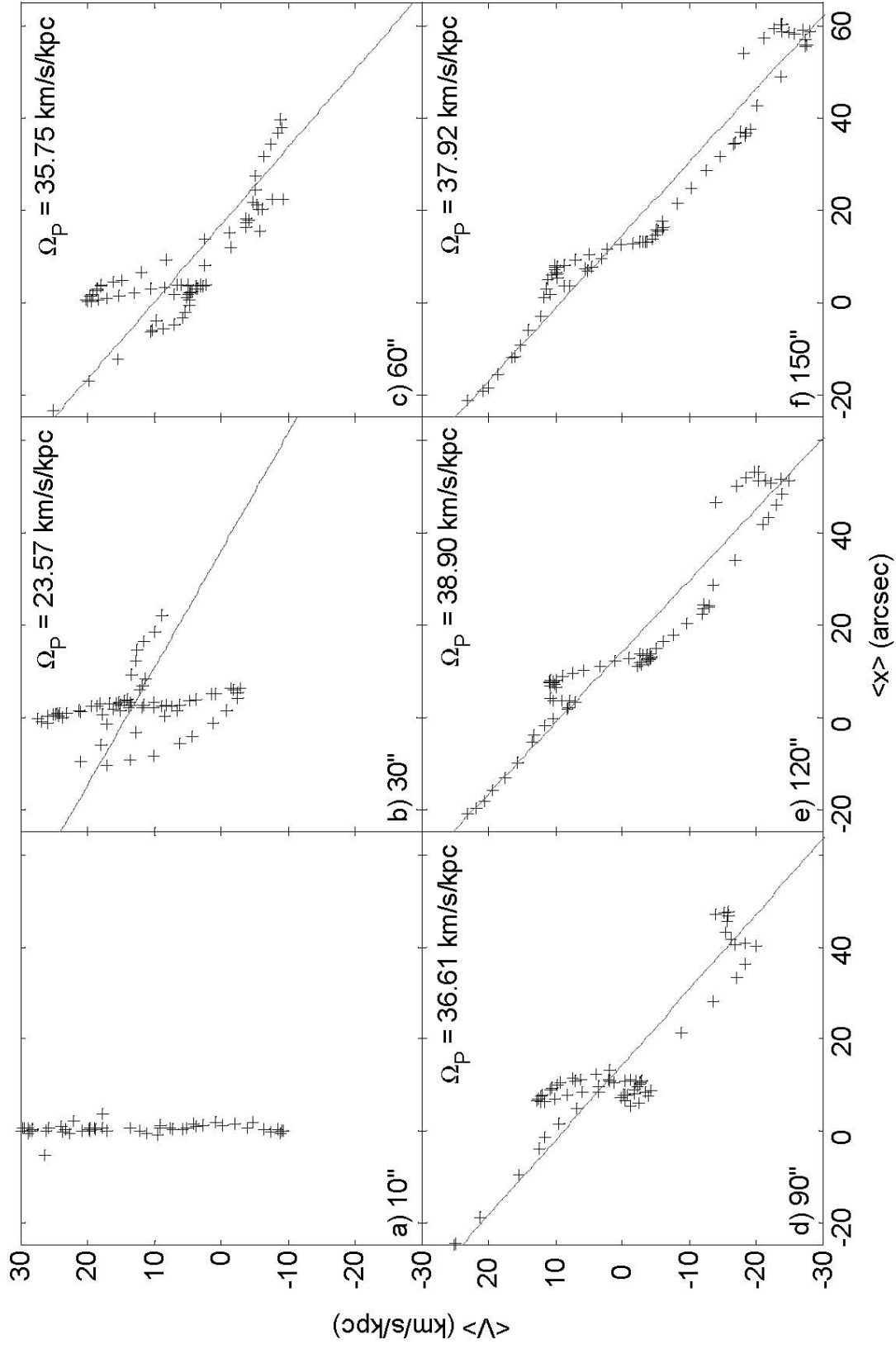

Figure 9: Convergence Test for apertures of M51. The six panels, a)-f), show the convergence of the TW method for M51 as the limits of integration for each aperture are extended outwards from the center of the galaxy. When the limits are only halfway to the edge of the map, a linear trend is already well established and the resulting pattern speed within our error budget.

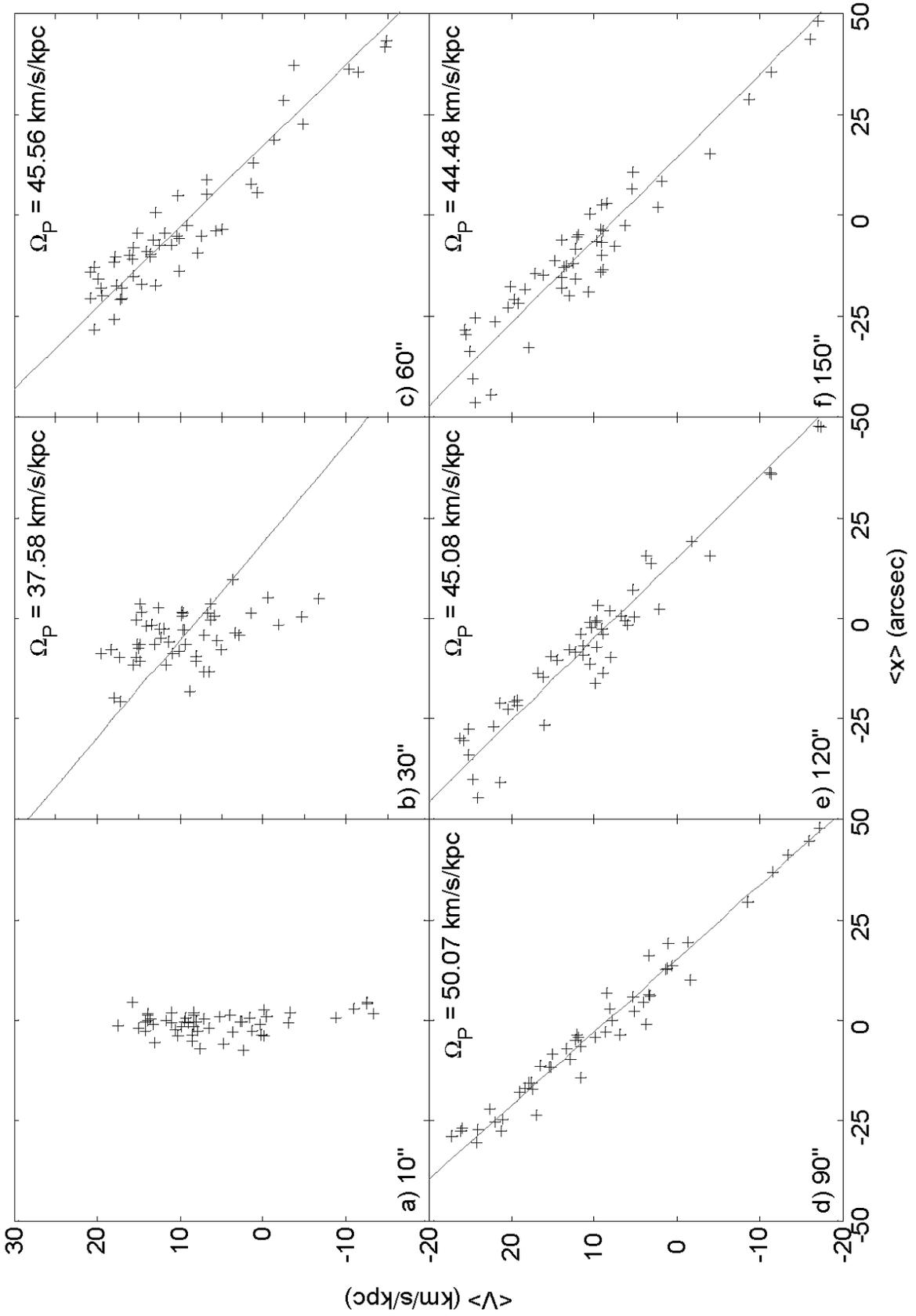

Figure 10: Same as Figure 9 except for M8

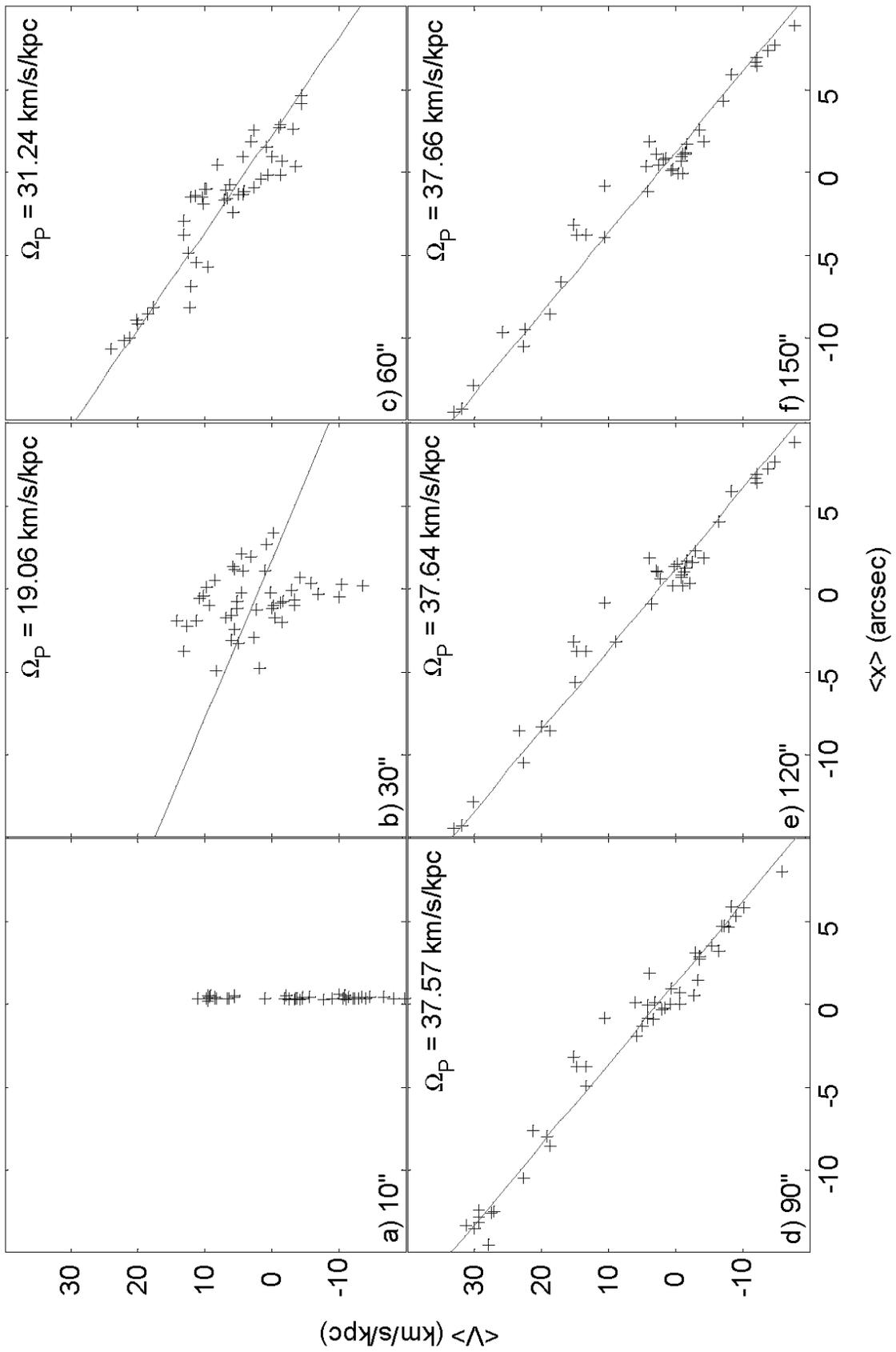

Figure 11: Same as Figure 9 except for NGC 6946.

## 5. CONCLUSIONS

The Tremaine-Weinberg method is a very powerful yet simple technique for determining the pattern speeds of spiral galaxies, as long as its assumptions are satisfied. We have used the TW method to measure the pattern speeds of M51, M83 and NGC 6946 and the resulting values are consistent with existing measurements using traditional resonance identification techniques where available. We have given arguments why molecular gas, traced by $^{12}$CO(1-0) emission, can be treated as pseudo-continuous for the purposes of the TW method, given a galaxy with a molecule dominated ISM. We have also tested the major sources of systematic errors in applying the TW method to CO emission and found their effects to be within TW's expected accuracy of 15%.

These results do not significantly constrain theories of spiral structure, though they are probably best interpreted in terms of the modal theory of Bertin and Lin (1989) because of our assumption that there is only one pattern speed in the region analyzed. Our results do not necessarily prove that no other speeds are present (and we have hints that another speed may be present in the bar of M51), but do suggest that one speed dominates. A broader study of a much larger sample of galaxies will lead to more solid constraints on spiral structure theories.

The TW method can be applied to many more galaxies with well-sampled, high resolution, full-flux maps of 21cm or CO emission, so long as either HI or $H_2$ dominates the ISM of the galaxy. Our tests of the sensitivity of the TW method to map resolution suggests that pattern speeds could be derived for a galaxies similar to M51 out to a distance of 25 Mpc using the Nobeyama 45-m telescope. Currently we are also exploring an idea suggested by M. Merrifield to apply the TW method to combined HI and $H_2$ maps to find $W_P$, which should make the TW method applicable to many more galaxies, and possibly even allow the CO®$H_2$ conversion factor to be constrained

ACKNOWLEDGEMENT: We would like to thank N. Nakai, S. García-Burillo, A. Andersson and W. Walsh for graciously supplying us with their CO maps. We would also like to thank M. Merrifield and our referee for their comments and feedback.